\documentclass[12pt]{amsart}
\usepackage{euler}
\usepackage{amsmath,amsthm,amsfonts,amscd}
\usepackage{latexsym}
\title[Projection methods]{Projection methods for some constrained systems}

\author[Pitanga]{Paulo Pitanga}
\address{P.P.\,-\,Instituto de Fisica, 
      Universidade Federal do Rio de Janeiro\\
       Caixa Postal 68528, Cidade Universit\'aria\\
        21945-970, Rio de Janeiro, RJ - Brazil}
\email {pitanga@if.ufrj.br} 

\author[Rodrigues]{Paulo R. Rodrigues}
\address{P.R.R.\,-\,Departamento de Geometria, Instituto de Matem\'atica\\ Universidade Federal Fluminense, 24020-140 Niter\'oi, RJ - Brazil}  
\email{\texttt{rodrigues@mat.uff.br} \\\texttt{rodriguespr@alternex.com.br}} 


\theoremstyle{plain}

\theoremstyle{definition}

\theoremstyle{remark} 
  
\newtheorem{remark}[]{Remark}  
 1  

\newcommand{\Ga}{\Gamma}

\newcommand{\w}{\mathbf{w}}

\def\Ri{{\rm I\kern-.25em\, R}}     

\begin{document} 
\begin{abstract}
This article is concerned with a geometric tool given by a pair of  projector operators defined by almost product structures  on finite dimensional manifolds, polarized by a distribution of constant rank and also endowed with some geometric structures (Riemann, resp. Poisson, resp. symplectic). The work is motivated by non-holonomic and sub-Riemannian geometry of mechanical systems on finite dimensional manifolds. Two examples are given.
\end{abstract}
  
\maketitle 
\section{Introduction}
\label{intro}\setcounter{page}{1}
In general, a classical constrained mechanical system consists in three basic ingredients: an $n$-dimensional configuration manifold $W$, a {\itshape polarization} ${\mathcal D}$ on $W$, which is to say a distribution ${\mathcal D}\colon z\in W \to {\mathcal D}_z \subset T_zW$ and an auxiliary geometric structure. ${\mathcal D}$ is hereafter supposed smooth, of constant rank (dim ${\mathcal D}_z=$ constant, for all $z$), and the Lie algebra of vector fields taken as sections of ${\mathcal D}$ over $W$ span the tangent space of $W$ at each point (thus ${\mathcal D}$ is non-integrable, i.e. non-involutive in the Frobenius sense).  We recall that a curve $\varphi(t)$ in $W$ satisfies the constraints (is polarized) if $\dot{\varphi}(t) \in {\mathcal D}_{\varphi(t)}$.

With these assumptions, we may outline some interesting directions on polarized systems: suppose first that the auxiliary geometric structure is given by a Riemannian metric tensor ${\mathbf g}$ such that $TW$ splits as $TW={\mathcal D}\oplus{\mathcal D}^{\perp}$, where ${\mathcal D}^{\perp}$ is the ${\mathbf g}$-orthogonal complement of ${\mathcal D}$ in $TW$: ${\mathbf g}(v,w)=0, v \in {\mathcal D}_z, w\in {\mathcal D}^{\perp}_z$, for all $z\in W$ (we use the same notation for the distribution and the corresponding sub-bundle of $TW$). Sometimes, the polarization ${\mathcal D}$ is said {\itshape horizontal}, and Vaisman \cite{IV-0}, \cite{IV-1}, following Reinhardt \cite{Rei}, coined the triple $(W,{\mathcal D},{\mathbf g})$ a {\itshape Riemannian almost foliated manifold}. If ${\mathcal D}$ is Frobenius integrable,  $(W,{\mathcal D},{\mathbf g})$ is said a {\itshape Riemannian foliated manifold}. 

In classical mechanics, the triple $(W,{\mathcal D},{\mathbf g})$ defines two different important mathematical structures: 
\begin{itemize}
\item[(1)] {\itshape Non-holonomic} (NH) mechanics by assuming that the trajectories satisfy D'Alem\-bert's principle of virtual work: the constraining force must be perpendicular to the horizontal subspace, since it does not produce work (see 
\S \ref{4.1}; see also ref. \cite{Hermann}, p. 85 or ref. \cite{NFUV}). 
\item[(2)] {\itshape Vakonomic} (VAK) mechanics by assuming that the trajectories do not obey D'Alem\-bert principle and satisfy a Lagrange variational principle (see Arnol'd et. al. \cite{Arnoldo} for further information).  
\end{itemize} 
Unless the distribution ${\mathcal D}$ is Frobenius integrable, VAK mechanics gives different geodesic equations from NH mechanics, and the comparison between these structures was ellucidated in ref. \cite{CMDM} (previous works on the subject are, for instance, references \cite{Cardin} and \cite{LMurray}). Recall from ref. \cite{vershik} that VAK mechanics is related to the so called sub-Riemannian (SR) or Car\-not-Ca\-ra\-theo\-do\-ry geometry (see the book edited by A. Bella\"{\i}che and J-J. Risler \cite{Bel}, R. Montgomery \cite{Montbook}, I. Kupka \cite{Kupka0} for further details in SR geometry, and Koiller et al. \cite{KRP1} for an example in sub-Riemannian Lagrange mechanics). 

Now, the splitting $TW={\mathcal D}\oplus{\mathcal D}^{\perp}$ means that $W$ admits an {\itshape almost product structure} {\bf a.p.s.}, for brevity, i.e,  $W$ is endowed with a tensor field $\boldsymbol{\Ga}$ of type $(1,1)$, taken as a vector valued one form on the tangent bundle $\boldsymbol{\Ga}\colon TW\to TW$, of involutive character, i.e, such that $\boldsymbol{\Ga}^2=\boldsymbol{id}$ (Schouten \cite{Schouten}, Nickerson-Spencer \cite{Nickerson} and Walker \cite{Walker}). In fact, there are defined the following bundle maps: a ${\mathcal D}^{\perp}$-valued one form ${\boldsymbol P}\colon TW\to {\mathcal D}^{\perp}$,  (a tensor field of type $(1,1)$ on $W$),  such that ${\mathbf P}$ is a projection operator onto ${\mathcal D}^{\perp}$: ${\boldsymbol P}\circ {\boldsymbol P}={\boldsymbol P}^2={\boldsymbol P}$, and a complementary ${\boldsymbol Q}=\boldsymbol{id}-{\boldsymbol P}\colon TW \to {\mathcal D}$. Then the pair $({\boldsymbol P},{\boldsymbol Q})$ defines $\boldsymbol \Ga={\boldsymbol Q} - {\boldsymbol P}$, with eigenvalues $1$ and $-1$, and associated eigenspaces ${\mathcal D}$ and ${\mathcal D}^{\perp}$, respectively. We remark that the interplay of {\bf a.p.s.} with covariant derivatives is one of the themes of Hermann's book \cite{Hermann} (see also \cite{deLR}).

On other hand, let us suppose that the polarized manifold $(W,{\mathcal D},{\mathbf g})$ is endowed with an {\bf a.p.s.} $\boldsymbol \Ga$, and in addition the eigenvector bundle corresponding to the eigenvalue $1$ is precisely ${\mathcal D}_z$, at each point $z$ of $W$. Then we have the projectors $${\boldsymbol Q}= (1/2)(\boldsymbol{id}+\boldsymbol{\Ga}), {\boldsymbol P}=(1/2)(\boldsymbol{id}-\boldsymbol{\Ga}),$$ 
and $T_zW = {\mathcal D}_z \oplus {\mathcal D}^{{\mathbf c}}_z, \,\forall \,z \in W$, where ${\mathcal D}^{{\mathbf c}}=\boldsymbol{im}\,{\boldsymbol P}$, image of ${\boldsymbol P}$ in $TW$. Therefore, an alternative case may be considered, where the subspaces ${\mathcal D}^{{\mathbf c}}_z\subset T_zW$ are hereafter supposed of complementary constant dimension,  everywhere transversal to those of ${\mathcal D}$, for all $z \in W$. For simplicity, we shall call this decomposition by  {\itshape oblique}.  
 
One more polarized situation is illustrated by Poisson manifolds. Consider the pair $(W,\boldsymbol \Pi)$, where ${\boldsymbol \Pi}$ is a twice contravariant skew-symmetric tensor field, verifying $[{\boldsymbol \Pi},{\boldsymbol \Pi}]=0$,  where $[\, ,\, ]$ are the Schouten brackets (see ref. \cite{MR1}, \cite{IV} for further details). The tensor ${\boldsymbol \Pi}$ is called {\it Poisson's tensor field} or {\it Poisson structure}, and $(W,\boldsymbol  \Pi)$ a Poisson manifold. The Poisson structure induces a bundle morphism ${\bf \sharp}: T^{\star} W \rightarrow TW$  such that $\beta \,{\bf \sharp} \, (\alpha) = {\boldsymbol \Pi}(\alpha, \beta)$, where $\alpha$ and $\beta$ are one forms on $W$. In particular, $dg\, {\bf \sharp}\,(df) = {\boldsymbol \Pi}(df, dg)= \{ f, g \}$ is the well known Poisson bracket of $f, \, g \in C^{\infty} (W)$,  the space of $C^{\infty}$ functions on $W$. 

Let us suppose that the Poisson bivectors are of constant rank $< n$. Then the {\itshape characteristic} distribution ${\mathcal S}\colon z\in W \mapsto {\mathbf \sharp}_z (T_{z}^{\star}W)$  is differentiable and completely integrable, and defines a foliation of $W$ such that each leaf ${\mathbf S}$ is endowed with a unique symplectic structure and the tangent space $T_z{\mathbf S}$ through each point $z$ is $\sharp_z(T^{\star}_zW)$. Therefore, we may also consider the case where the {\bf a.p.s.} is defined by a polarization $(W,{\mathcal S},{\boldsymbol \Pi})$, such that $T_zW = T_z{\mathbf S}\oplus {\mathbf S}^{{\mathbf c}}_z$, with the second factor being a complementary distribution of constant dimension. In fact, these comments motivate a study of the inter-relation of almost product geometry with some fields of classical mechanics.  

Indeed, the purpose of this article is to retake this subject for the cases where the projectors can be modeled by a Riemannian or a  Poisson (resp. symplectic) structure on a finite dimensional manifold. This means that we will be in the context of the determination of appropriate projectors defined by these geometric objects. We propose as a first task to re-examine the relationship which exists between the almost product geometry with a (non-degenerate) Riemannian structure. We adopt the Pfaffian view point of ref. \cite{Montbook}, and we assume that $(W,{\mathbf g})$, endowed with a set of $n-m$-linearly independent one-forms $w^{\alpha}$, such that $w^{\alpha}=0$ defines a non-involutive distribution of constant dimension, with a complementary integrable distribution, i.e, $W$ is a foliated Riemannian manifold. 
We shall express the metric in terms of a local coframe 
$\{dz^{a},\omega^{\alpha}\}$ on the configuration manifold. Therefore $W$ admits an "oblique {\bf a.p.s.}". Naturally we may  orthogonalize the {\bf a.p.s.}, i.e,  the {\bf a.p.s.} is  defined by a cobasis of orthogonal covectors with respect to the given metric. However, as we shall see in the examples, we may work directly with the oblique situation to obtain the projected dynamical equations, avoiding the application of the Gram-Schmidt procedure. This will be not only convenient for matrix calculations, but also to set up other studies like the equivalence problem (ref. \cite{Montbook}, \cite{KRP}).  Particular cases are the sub-Riemannian geodesic problem and the Vakonomic variational approach.

Next, as a second task, we consider an involutive polarization on a Poisson manifold, defined by its integrable symplectic foliation.  We apply the projector method to obtain the Poisson structure for the transverse manifold of the symplectic leaves.  This could be seen as an application of the previous study in the sense that the metric is replaced by a Poisson structure, and the integrable distribution is the characteristic distribution. We profit the occasion to give the Dirac formula for constraint manifolds in the transverse situation (see ref. \cite{MR1} for the symplectic submanifold case; the technique used is the usual, but we have searched the literature and have not found it for the transverse situation). The reader will find more about Dirac mechanics in non-holonomic contexts in references \cite{Cantrijn} and \cite{Cantrijn2}.

Finally, we would like to stress here that the projector method is an adequate tool to treat some variational problems in which the extremal curves as well as the comparison curves (associated with a given Lagrangian function) are required to fulfill conditional equations (the constraints). As it is well known, the oldest problem of this type was solved by Pappus in the third century A.D. These variational problems are called  {\itshape The Problem of Lagrange}, who  first formulated the problem clearly \cite {Ca}. The projector method is based on the (orthogonal) decomposition of the virtual displacement, which make it right for dealing with a generalized form of D'Alembert's Principle from which the equation of motion of mechanical systems are usually derived (see ref. \cite {Da} and also the  book of Arnol'd, ref. \cite{ArnoldMec}, p. $91 - 95$). We remark that, when the  method is applied to the Problem of Lagrange,  neither {\itshape Lagrange's multipliers} nor  {\itshape elimination of coordinates} is required to obtain the equations of motion. In this way, this approach, is appropriate to implement the canonical quantization of constrained systems, because the ambiguities introduced by the Lagrange's multipliers are eliminated. It is also appropriate for setting up computer calculations  for large multibody systems which appears in control problem of mechanical systems and robotics (for example, see ref. \cite{robotics} for a computer use of {\itshape IEEE Scheme Programming} in Mechanics).  

This paper is structured in three sections. In section $2$ we examine some intrinsic and local properties of the projector method on a Riemannian manifold. We begin with some intrinsic considerations using the so-called {\it musical} bundle morphisms $\sharp_{\mathbf g}, \, \flat_{\mathbf g}$, induced by the Riemannian metric ${\mathbf g}$. Next we suppose that ${\mathcal D}$ is integrable, characterized by a set of $k$-linearly independent one-forms. We express the pair of bundle projectors in terms of this set, we obtain the local expressions for the corresponding bundle projection morphisms, and then we give a local description in terms of the Riemannian metric.  

In section $3$, we replace the geometric structure ${\mathbf g}$ by ${\boldsymbol \Pi}$, and we study the role of the bundle projectors to obtain the Poisson structure for the transverse manifold of the symplectic leaves. We shall return to this situation in the second example of the last section $4$. Indeed, this section is only devoted to applications of the method. We start subsection $4.1$ with a brief review on D'Alembert's Principle and then we study the so-called Chaplygin-Caratheodory sleigh, a prototype of a non-holonomic constrained system. Next, in subsection $4.2$, we examine a system consisting of a free particle in $\Ri^3$, subjected to the non-holonomic contact form $w=dz-ydx$. From the projector's viewpoint one obtains a very well known non-holonomic Lie algebra, the Heisenberg algebra, a fundamental example in sub-Riemannian geometry. We conclude the example with a study of the motion of the particle in the phase space using the underlying Poisson structure. 

The following convention will be adopted: capital roman letters $I,J,K,$ etc. run from $1$ to $n$. Lower case roman characters $a,b,c $ run from $1$ to $m$, representing the constraint distribution. Greek characters $\alpha, \beta, \gamma$, etc., run from  $1$ to $n-m$.  Summation over repeated indices is assumed unless otherwise stated. By a differentiable manifold, we shall mean ${\mathcal C}^{\infty}$, connected, separable and Hausdorff.
  
\section{Polarized Riemannian manifolds}\label{Rcase}
\subsection{An intrinsic relation}
Let $(W,{\mathcal D},{\mathbf g})$ be a $n$-dimensional Riemannian  manifold, supposed endowed with an {\bf a.p.s.} $\boldsymbol \Ga$, compatible with ${\mathcal D}$ in the sense that the eigenvector bundle corresponding to the eigenvalue $-\,1$ is precisely ${\mathcal D}_z$, at each point $z$ of $W$.
Let us take the corresponding bundle projections ${\boldsymbol P}, \, {\boldsymbol Q}={\boldsymbol{id}}-{\boldsymbol P}$ so that ${\boldsymbol Q}\colon TW\to {\mathcal D}$ projects onto ${\mathcal D}$, and ${\boldsymbol P}\colon TW\to {\mathcal D}^{{\mathbf c}}$ projects onto a complementary  distribution ${\mathcal D}^{{\mathbf c}}$.

Now, the tensor ${\mathbf g}$ defines a bundle isomorphism $\sharp_{\mathbf g}\colon T^{\star}W\to TW$ with inverse denoted by $\flat_{\mathbf g}\colon TW\to T^{\star}W$ (the so-called {\itshape musical} morphisms), defined respectively by
\[
\sharp_{\mathbf g}(\phi)= \mathbf g_{\star}(\phi,\bullet)\stackrel{\mathrm{def}}{=}Z_{\phi}, \,\, \flat_{\mathbf g}(Z)=\mathbf g^{\star}(\bullet,Z)\stackrel{\mathrm{def}}{=}{\phi}_Z.
\]
As ${\mathbf g}$ is symmetric, one has $\sharp_{\mathbf g}=\sharp_{\mathbf g}^{\star}$, where $\sharp_{\mathbf g}^{\star}\colon T^{\star}W\to TW$ is the adjoint operator $\sharp_{\mathbf g}^{\star}(\phi)=\phi \circ \sharp_{\mathbf g}$.
So, if we set $\sharp_{\boldsymbol Q}= {\boldsymbol Q}\circ \sharp_{\mathbf g}, \,\, \sharp_{\boldsymbol P}= {\boldsymbol P} \circ \sharp_{\mathbf g}$, then
\begin{eqnarray*}
\sharp_{\boldsymbol Q}&=&{\boldsymbol Q}\circ \sharp_{\mathbf g} = ({\boldsymbol Q}^{\star})^{\star}\circ \sharp_{\mathbf g} =\sharp_{\mathbf g}\circ {\boldsymbol Q}^{\star}=\sharp_{\boldsymbol Q}^{\star}\\
\sharp_{\boldsymbol P}&=&{\boldsymbol P}\circ \sharp_{\mathbf g} = ({\boldsymbol P}^{\star})^{\star}\circ \sharp_{\mathbf g} =\sharp_{\mathbf g}\circ {\boldsymbol P}^{\star}=\sharp_{\boldsymbol P}^{\star},
\end{eqnarray*}
and it follows that
\begin{eqnarray}
\mathbf g_{\star}({\boldsymbol Q}^{\star}(\phi),\psi)&=&\psi\,(\sharp_{\mathbf g}\circ{\boldsymbol Q}^{\star})\,\phi=\psi\,({\boldsymbol Q}\circ\sharp_{\mathbf g})\,\phi ={\boldsymbol Q}^{\star}(\psi)\sharp_{\mathbf g}\,\phi=
\mathbf g_{\star}(\phi,{\boldsymbol Q}^{\star}(\psi))\\
\mathbf g_{\star}({\boldsymbol Q}^{\star}(\phi),\psi)&=&\psi\,(\sharp_{\mathbf g}\circ{\boldsymbol Q}^{\star})\,\phi=\psi\,(\sharp_{\mathbf g}\circ({\boldsymbol Q}^{\star})^{2})\,\phi=\psi\,({\boldsymbol Q}\circ\sharp_{\mathbf g}\circ{\boldsymbol Q}^{\star})\,\phi\nonumber\\
&=&{\boldsymbol Q}^{\star}(\psi)\,\sharp_{\mathbf g}\,{\boldsymbol Q}^{\star}(\phi) =\mathbf g_{\star}({\boldsymbol Q}^{\star}(\phi),{\boldsymbol Q}^{\star}(\psi)). 
\end{eqnarray}
Also,
\begin{eqnarray}
\mathbf g_{\star}({\boldsymbol P}^{\star}(\phi),{\boldsymbol P}^{\star}(\psi))&=&\mathbf g_{\star}({\mathbf I}(\phi)-{\boldsymbol Q}^{\star}(\phi),{\mathbf I}(\psi)-{\boldsymbol Q}^{\star}(\phi))\nonumber\\
&=&{\mathbf g}_{\star}(\phi,\psi)- 2\,\mathbf g_{\star}({\boldsymbol Q}^{\star}(\phi),\psi)+\mathbf g_{\star}({\boldsymbol Q}^{\star}(\phi),{\boldsymbol Q}^{\star}(\psi))\nonumber\\
&=&
{\mathbf g}_{\star}(\phi,\psi)- \mathbf g_{\star}({\boldsymbol Q}^{\star}(\phi),{\boldsymbol Q}^{\star}(\psi)).
\end{eqnarray}
If we set $$\mathbf g_{{\boldsymbol Q}}=\mathbf g_{\star}({\boldsymbol Q}^{\star}(\bullet),{\boldsymbol Q}^{\star}(\bullet)),\,\,\,\mathbf g_{{\boldsymbol P}}=\mathbf g_{\star}({\boldsymbol P}^{\star}(\bullet),{\boldsymbol P}^{\star}(\bullet))$$
then 
$\mathbf g_{\star} = \mathbf g_{{\boldsymbol P}}+\mathbf g_{{\boldsymbol Q}}$,
as it would be expected, i.e. the almost product structure ${\boldsymbol \Gamma}^{\star} = {\boldsymbol P}^{\star}-{\boldsymbol Q}^{\star}$ is such that 
${\mathbf g_{\star}}({\boldsymbol \Gamma}^{\star} (\phi),{\boldsymbol \Gamma}^{\star} (\psi))$ $=$ ${\mathbf g_{\star}}(\phi,\psi)$ for all  $\phi$, $\psi$. Obviously,  \begin{equation}\label{obviou}
\sharp_{\mathbf g}=\sharp_{\boldsymbol Q} + \sharp_{\boldsymbol P}.
\end{equation}

\begin{remark} 
We observe that if we replace the tensor ${\mathbf g}$ by a Poisson tensor ${\boldsymbol \Pi}$ or by a symplectic structure ${\boldsymbol \Omega}$ then we have a similar result if the above compatibility assumption on ${\boldsymbol \Ga}$ is assumed (see also p. \pageref{transverse Poisson}). $\quad \Box$ 
\end{remark}

\subsection{Local expressions for the case of a foliated manifold} Let us suppose now that $\mathcal{D}$ is a completely integrable distribution on $W$, of constant rank $n-m$, and so $(W,{\mathcal D},{\mathbf g}, {\boldsymbol \Ga})$ is a Riemannian foliated almost product manifold. The distribution defines a foliation on $W$, denoted also by the same symbol ${\mathcal D}$, to simplify things. The tangent bundle $T{\mathcal D}$  is the vector sub-bundle of $TW$ such that ${\mathcal D}_z=T_z{\mathcal E}_z$ for any leaf ${\mathcal E}$ of ${\mathcal D}$ and any $z\in {\mathcal E}$. 

Let us consider the splitting  $T_zW = {\mathcal D}_z \oplus {\mathcal D}^{{\mathbf c}}_z, \,z \in W$, where ${\mathcal D}^{{\mathbf c}}$ is a distribution of subspaces ${\mathcal D}^{{\mathbf c}}_z\subset T_zW$, of complementary constant dimension, everywhere transversal to those of ${\mathcal D}$.  
To look for the local expressions, let $U$ be a neighborhood of $z\in W$, so that (due to the integrability of ${\mathcal D}$) the leaf is locally given by equations $z^a\equiv 0$. Thus we write ${\mathcal D}_z= \textrm{span}\{Y_\alpha=\partial/\partial z^\alpha\}$. Let $Y_{a}$ be a set of linearly independent vectors with ${\mathcal D}^{{\mathbf c}}_z= \textrm{span}\{Y_{a}\}$. Furthermore, writing $Y_{a}=\Gamma^I_{a}\,({\partial/\partial z^I})$ in the coordinate basis one obtains $ {\partial/\partial z^{a}}\,\, - \Gamma^{\alpha}_{a}\,{\partial/\partial z^{\alpha}},$
as a new basis, for suitable functions $\Gamma^{\alpha}_{a}(z)$ on $W$. Then we define the projectors ${\boldsymbol P}\colon T_zW\to {\mathcal D}_z^{{\mathbf c}},\, {\boldsymbol Q}={\boldsymbol{id}}-{\boldsymbol P}\colon T_zW\to {\mathcal D}_z$:
\begin{eqnarray}\label{eq1-1}
{\boldsymbol P}(Z)= Z^{a}\,(\frac{\partial}{\partial z^{a}}- \Gamma^\alpha_{a}\,\frac{\partial}{\partial z^{\alpha}})
\,,\,\,\,{\boldsymbol Q}(Z)&=& (Z^{\alpha} + \Gamma^\alpha_{a}\,Z^{a})\,\frac{\partial}{\partial z^{\alpha}},
\end{eqnarray}
with $Z=Z^{a}({\partial}/{\partial z^{a}}) + Z^{\alpha}({\partial}/{\partial z^{\alpha}})$. In matrix notation:
\begin{alignat}{2}\label{mproj1-v}
{\boldsymbol P}=
\left(
\begin{matrix}
      {\rm id} &0 \\
    -\Gamma^\alpha_{a} &0
\end{matrix}
\right),
\qquad
&&
{\boldsymbol Q}=
\left(
\begin{matrix}
       0 &0 \\
   {\Gamma^\alpha_{a}} &{\rm id}
\end{matrix}
\right).
\end{alignat}
To simplify the notation we set heretofore $X_{a}= {\partial/\partial z^{a}}\,\, - \Gamma^{\alpha}_{a}\,{\partial/\partial z^{\alpha}}$. The following figure illustrates the situation:
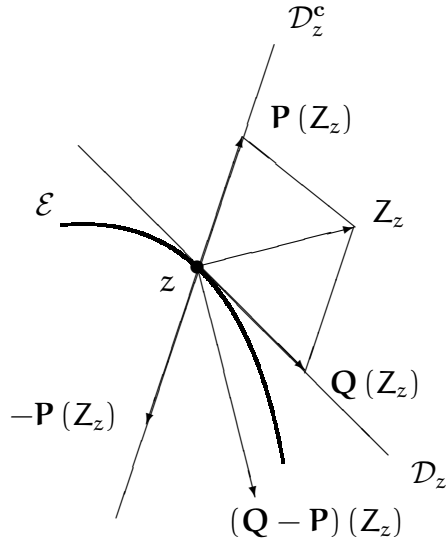
\begin{center}
\unitlength=.60mm
\linethickness{1pt}
\begin{picture}(95.00,138.00)
\bezier{388}(13.67,82.67)(53.00,86.33)(62.67,30.00)
\put(44.00,73.67){\vector(4,1){34.33}}
\put(44.00,74.00){\vector(1,-1){23.50}}
\put(43.67,73.67){\vector(1,-4){12.80}}
\put(43.89,73.00){\vector(-1,-3){11.50}}
\linethickness{0.4pt}
\put(17.33,100.33){\line(1,-1){68.67}}
\put(78.67,82.33){\line(-5,4){25.00}}
\put(60.67,122.67){\line(-1,-3){35.00}}
\put(78.33,82.33){\line(-1,-3){10.80}}
\put(37.00,69.67){\makebox(0,0)[cc]{$z$}}
\put(86.00,85.33){\makebox(0,0)[cc]{$Z_z$}}
\put(83.00,47.33){\makebox(0,0)[cc]{${\boldsymbol Q}\,(Z_z)$}}
\put(95.00,27.33){\makebox(0,0)[cc]{${\mathcal D}_z$}}
\put(69.00,106.00){\makebox(0,0)[cc]{${\boldsymbol P}\,(Z_z)$}}
\put(10,87.33){\makebox(0,0)[cc]{${\mathcal E}$}}
\put(68.00,128.00){\makebox(0,0)[cc]{${\mathcal D}^{{\mathbf c}}_z$}}
\put(70.00,16.50){\makebox(0,0)[cc]{${(\boldsymbol Q - \boldsymbol P)}\,(Z_z)$}}
\put(14,40){\makebox(0,0)[cc]{${-\boldsymbol P}\,(Z_z)$}}
\put(44.00,73.33){\vector(1,3){9.67}}
\put(43.67,73.33){\circle*{2.98}}
\end{picture}
\end{center}

Let $w^{\alpha}=dz^{\alpha} + \Gamma^{\alpha}_{a}(z) dz^{a}$  be the set of independent $1$-forms such that $\{dz^{a},w^{\alpha}\}$ is the corresponding cobasis for the cotangent space $T^*_zW$. 
Then 
$$
T^{\star}_z{W}=({\mathcal D}^{{\mathbf c}}_z)^{\star} \oplus ({\mathcal D}_z)^{\star} = \textrm{span}\{dz^{a}\} \oplus \textrm{span}\{w^{\alpha}\},
$$
and we may set 
$$
{\boldsymbol P}=dz^{a}\otimes X_{a},\,\,{\boldsymbol Q}=w^{\alpha}\otimes Y_{\alpha}.
$$

We observe that locally (see (\ref{obviou} and (\ref{eq1-1})),
\begin{eqnarray*}
\sharp_{\boldsymbol P}(\phi)&=&{\boldsymbol P}\circ \sharp_{\mathbf g}(\phi)=Z^{a}_{\phi}(\frac{\partial}{\partial z^{a}}- \Gamma^{\alpha}_{a}\,\frac{\partial}{\partial z^{\alpha}})\\
\sharp_{\boldsymbol Q}(\phi)&=&{\boldsymbol Q}\circ \sharp_{\mathbf g}(\phi)=(Z^{\alpha}_{\phi} + \Gamma^{\alpha}_{a}\,Z^{a}_{\phi})\,\frac{\partial}{\partial z^{\alpha}}\,.
\end{eqnarray*}

Next we shall study these projectors in terms of the Riemann matric 
${\mathbf g}=g_{{I}{J}}dz^I\otimes dz^J$.  
To do this we recall that ${\mathbf g}$ has the following expressions (see ref. \cite{IV-0}): in the basis $\{dz^{a},w^{\alpha}\}$ 
\begin{eqnarray}\label{met-ger-1}
&&{\mathbf g^{\star}}=(g_{ab}-2\,g_{a{\alpha}}\,\Gamma^{\alpha}_{b}+g_{\alpha\beta}\,
\Gamma^{\alpha}_{a}\,\Gamma^{\beta}_{b})dz^{a}\otimes dz^{b}\\&& + 2\,(g_{a\alpha} - g_{\alpha\beta}\,\Gamma^{\beta}_{a})\, dz^{a}\otimes w^{\alpha} + g_{\alpha\beta}\, w^{\alpha}\otimes w^{\beta}\nonumber\\
&&\stackrel{\textrm{def}}{=} F_{ab}\, dz^{a}\otimes dz^{b} + F_{a\alpha} \,dz^{a}\otimes w^{\alpha} + G_{\alpha\beta}\, w^{\alpha}\otimes w^{\beta} \nonumber 
\end{eqnarray}
and in the basis $\{X_a, Y_{\alpha}\}$,  
\begin{eqnarray*}
&&{\mathbf g_{\star}} =  g^{ab}\, X_{a} \otimes X_{b} + 2\,(g^{a\alpha} + g^{ab}\,\Gamma^{\alpha}_{b}) \, X_{a} \otimes Y_{\alpha}\nonumber\\&& + (g^{\alpha \beta} + 2 g^{a\alpha}\,\Gamma^{\beta}_{a} + g^{ab}\,\Gamma^{\alpha}_{a}\,\Gamma^{\beta}_{b})\, Y_{\alpha}\otimes Y_{\beta}  \nonumber\\
&&\,\,\stackrel{\textrm{def}}{=} G^{ab}\, X_{a} \otimes X_{b} + G^{a\alpha}\, X_{a} \otimes Y_{\alpha} + G^{\alpha \beta}\,Y_{\alpha}\otimes Y_{\beta}. 
\end{eqnarray*}

Let $\xi^{\alpha} = {\mathbf g_{\star}}(w^{\alpha},\bullet)$. Then
$w^{\beta}(\xi^{\alpha}) = {\mathbf g_{\star}}(w^{\alpha},w^{\beta})= G^{\alpha\beta}$. As the matrix with entries $G^{\alpha\beta}$ is invertible, let $G_{\alpha\beta}$ be the set of functions which are the entries of the inverse. We define the tensor
\begin{eqnarray}\label{proj-M}
{\mathbf q}= {\mathbf g^{\star}}(\bullet,\xi^{\alpha})\,G_{\alpha\beta}\,{\mathbf g_{\star}}(w^{\beta},\bullet)\,.
\end{eqnarray}
Then one obtains
\begin{equation}\label{comp9}
{\mathbf q} = G_{\alpha\beta}\,w^{\alpha}\otimes \xi^{\beta},
\end{equation}
and so ${\mathbf q}(\xi^{\beta})=(G^{\beta \alpha}\,G_{\alpha \gamma})\,\xi^{\gamma}$. As
\begin{eqnarray*}
{\mathbf q}^{2}&=&[{\mathbf g^{\star}}(\bullet,\xi^{\alpha})\,G_{\alpha\beta}\,{\mathbf g_{\star}}(w^{\beta},\bullet)]\,[{\mathbf g^{\star}}(\bullet,\xi^{\gamma})\,G_{\gamma\theta}\,{\mathbf g_{\star}}(w^{\theta},\bullet)]\\
&=&{\mathbf g^{\star}}(\bullet,\xi^{\alpha})\,G_{\alpha\beta}\,[{\mathbf g_{\star}}(w^{\beta},\bullet)\,{\mathbf g^{\star}}(\bullet,\xi^{\gamma})]\,G_{\gamma\theta}\,{\mathbf g_{\star}}(w^{\theta},\bullet)\\
&=&{\mathbf g^{\star}}(\bullet,\xi^{\alpha})\,G_{\alpha\beta}\,G^{\beta\gamma}\,G_{\gamma\theta}\,{\mathbf g_{\star}}(w^{\theta},\bullet)={\mathbf q},
\end{eqnarray*}
${\mathbf q}$ is a projector onto the space spanned by the $\xi^{\alpha}$'s, with complementary projector ${\mathbf p}$. 

Let us suppose now that $g_{a \alpha} - g_{\alpha\beta}\,\Gamma^{\beta}_{a}$ vanishes in (\ref{met-ger-1}). Then ${\mathcal D}_z$ and ${\mathcal D}^{{\mathbf c}}_z$ are ${\mathbf g}$-orthogonal, ${\mathcal D}^{{\mathbf c}}={\mathcal D}^{{\perp}}$, ${\mathbf g^{\star}}(X_{a},Y_{\alpha})=0$ and so ${\mathbf g^{\star}}$ admits the following diagonal form with respect to $\{dz^{\alpha}, w^a\}$,
\[
{\mathbf g^{\star}}=G_{{a}{b}}\, dz^{a}\otimes dz^{b} + G_{\alpha\beta}\, w^{\alpha}\otimes w^{\beta},\,\,G_{a b}={\mathbf g^{\star}}(X_{a},X_{b}),\,G_{\alpha\beta}=g^{\star}(Y_{\alpha},Y_{\beta})
\]
(or ${\mathbf g_{\star}}=G^{{a}{b}}\,X_{a}\otimes X_{b} + G^{\alpha\beta}\, Y_{\alpha}\otimes Y_{\beta}$,
where $G^{a b}$, resp., $G^{\alpha\beta}$ are the entries of the inverse matrix of $(G_{a b}$, resp., $(G_{\alpha\beta})$). As $\xi^{\alpha} = {\mathbf g_{\star}}(w^{\alpha},\bullet)=G^{\alpha \gamma}\, Y_{\gamma}$, then it is easily verified that 
\begin{equation}\label{defproj-1}
{\boldsymbol q}=w^{\alpha}\otimes Y_{\alpha}={\boldsymbol Q}
\end{equation}
and so  $\textrm{Im}\,{\boldsymbol q}=\textrm{span}\{Y_{\alpha}\}$ (and obviously ${\boldsymbol p}={\mathbf P}$), i.e, 
these projectors have the same local matrix expression given by (\ref{mproj1-v}). 
\section{Projectors and Transverse Poisson structures}
In this section we first consider a polarization on a Poisson manifold, defined by its integrable symplectic foliation, and the local expression of the corresponding projectors in terms of the symplectic form. In the second part we study the role of the bundle projector $TW\vert M \stackrel{{\mathbf p}_{TM}}{\rightarrow} TM$, where $M\subset W$ is a given transversal (holonomic) manifold  of codimension $=$ dimension of the symplectic leaf  (see the Introduction), in the process of reduction of a Poisson manifold $(W,{\boldsymbol \Pi}_W)$. We use the projector method to obtain the local expression of the corresponding induced Poisson structure $(M,{\boldsymbol \Pi}_M)$.

Throughout this section we follow the following convention: local coordinates are now denoted by $z=(z^a, z^{u})$, the characters $a,b,c$ running from $1$ to $m$ and $u,v$ from $1$ to $n-m=k$. Greek characters $\alpha, \beta, \gamma$, etc., are used for differential forms on manifolds.

\subsection{Projectors for the symplectic foliation}
We assume that the smooth manifold $W$ is endowed with a Poisson structure - hereafter denoted by ${\boldsymbol\Pi}_W$ - of constant rank $m <\,n$, which induces a bundle morphism 
${\bf \sharp}\colon T^{\star} W \rightarrow TW$ such that $\beta \,{\bf \sharp} \, (\alpha) = {\boldsymbol\Pi}_W(\alpha, \beta)$, where $\alpha$ and $\beta$ are one forms on $W$. 

Let ${\mathbf S}$ be the unique symplectic leaf of the characteristic distribution $z\mapsto {\bf \sharp}_z(T^{\star}_zW)$ going through $z \in W$. Thus ${\mathbf S}$ is obviously a Poisson submanifold. Now, as the rank of ${\boldsymbol\Pi}_W$ is constant, the symplectic leaves are of constant dimension $=m$,  and we may choose a decomposition $$T_zW = T_z{\mathbf S}\oplus {\mathbf S}^{{\mathbf c}}_z,$$with  $T_z{\mathbf S}= \textrm{span}\{Y_a\}$, where $Y_a=\partial/\partial z^a$, and ${\mathbf S}^{{\mathbf c}}_z= \textrm{span}\{X_{u}= \partial/\partial z^{u} - \sum_{a=1}^k\,A^a_{u}\,Y_a\}$, for a local coordinate chart $(U,(z^a,z^{u}))$. The Poisson bivector is locally
$
{\boldsymbol \Pi}_W=\frac{1}{2}\, \pi^{ab}\, Y_a \wedge Y_b$. 

So, from this local expression one obtains  an almost symplectic form ${\mathbf \Theta}$ (see Vaisman \cite{IV}, p.~37)
expressed as
$
{\mathbf \Theta}= \frac{1}{2}\, \lambda_{ab}\, w^a \wedge w^b$, 
with $w^a$ being the dual form of $Y_a$, and $\pi^{ba}\,\lambda_{ac}=\delta_c^b$ (recall that $\pi^{ab}=-\lambda^{ab}$). Therefore one obtains the following projectors : the first, taking into account that $${\mathbf \Theta}(\bullet,Y^a)=\lambda_{cb}\,w^c(Y_a)w^b=\lambda_{cb}\,
\delta^c_a\,w^b=\lambda_{ab}\,w^b,$$
is given by (compare with (\ref{comp9}))
\begin{eqnarray}\label{projsym-1-2}
{\mathbf \Theta}(\bullet,Y^a)\,\lambda_{ab}\,{\boldsymbol \Pi}_W(w^b,\bullet)&=&{\mathbf \Theta}(\bullet,Y^a)\,\,\lambda_{ab}\,\xi^b= {\mathbf \Theta}(\bullet,Y^a)\lambda_{ab}\,\pi^{bc}\,Y_c\nonumber\\
&=&{\mathbf \Theta}(\bullet,Y^a)\,\delta^c_a\,Y_c
={\mathbf \Theta}(\bullet,Y^a)\,Y_a\nonumber\\
&=&{\lambda_{ab}\,w^b\otimes Y_a}={\mathbf \Theta}(\bullet,Y^a)\,\lambda_{ab}\,{\boldsymbol \Pi}_W(dz^b,\bullet),
\end{eqnarray} 
and the second (compare with (\ref{defproj-1})), 
\begin{eqnarray}\label{projsym-1}
{\mathbf \Theta}(\bullet,\xi^a)\,\lambda_{ab}\,{\boldsymbol \Pi}_W(w^b,\bullet)&=&w^a\,\,\lambda_{ab}\,\xi^b
= \lambda_{ab}\,\pi^{bc}\,w^a \otimes Y_c\nonumber\\
&=&w^a\otimes Y_a={\mathbf \Theta}(\bullet,\xi^a)\,\lambda_{ab}\,{\boldsymbol \Pi}_W(dz^b,\bullet).
\end{eqnarray}

\subsection{The transverse holonomic case}
Let us suppose now that $M$ is a submanifold of the polarized Poisson manifold $(W,{\mathcal S},{\boldsymbol \Pi}_W)$, and denote by $$\mathsf{Ann}\,T_zM= \{ \alpha_{z} \in T^{\star}_{z}W; \alpha_{z} (T_zM) = 0\},$$the annihilator of $TM\subset TW\vert_M$ in $T^{\star}W$.  Suppose that
\[
\boldsymbol{(a)} \,\, T_zM \cap \sharp_z\,\mathsf{Ann}\,T_zM = \{0\}, \quad \boldsymbol{(b)} \,\, {\ker}\, \sharp_z \cap \mathsf{Ann}\,T_zM = \{0\},\; \forall z \in M
\]
which are equivalent to the condition $T_zW \equiv T_zM \oplus \sharp_z \, \mathsf{Ann}\,T_zM$ (see ref. \cite{LichneM}, p.~126, for a proof). Then Weinstein  (\cite{W}, p.~529) proved that $M$ is endowed with a Poisson structure ${\boldsymbol \Pi}_M$, defined by the composition 
\begin{equation}\label{DIAG}
T^{\star}M \stackrel{{\mathbf p}^{\star}_{TM}}{\longrightarrow} T^{\star}W\vert M \stackrel{\sharp}{\longrightarrow} TW\vert M \stackrel{{\mathbf p}_{TM}}{\longrightarrow} TM  
\end{equation}
where ${\mathbf p}_{TM}$ is the bundle projection along $\sharp_z (\mathsf{Ann}\,T_zM)$ onto $TM$ and ${\mathbf p}^{\star}_{TM}$ is its adjoint (Proposition 1.4 of ref. \cite{W}, p. 529/530). 
Now, it can be shown that (see ref. \cite{IV}, p. 39) 
\begin{itemize}
\item[-] the assumption $\boldsymbol{(a)}$ is equivalent to the statement that $T_zM\cap T_z{\mathbf S}$ is a symplectic subspace of $T_z{\mathbf S}$, and
\item[-] the assumption $\boldsymbol{(b)}$ is equivalent to the statement that $M$ is transversal to the symplectic leaf ${\mathbf S}$ passing through $z$. 
\end{itemize}
Therefore $T_{{z}}W \equiv T_{{z}}M + T_z{\mathbf S} $, and: \begin{itemize}
\item[-]the Poisson tensor of $W$ is the product of the Poisson tensor ${\boldsymbol \Pi}_{{\mathbf S}}$ by ${\boldsymbol \Pi}_{M}$, the so-called {\itshape transverse Poisson structure to ${\mathbf S}$} at $z$ (${\boldsymbol \Pi}_{{\mathbf S}}$ is induced by the symplectic structure of ${\mathbf S}$), 
\item[-] the transverse Poisson structure  of $M$  may be computed via the Dirac's bracket formula if further assumptions are made (see refs. \cite{JMR}, \cite{MR1}, Prop. 8.5.1, p. 226, \cite{OH}, Prop. 2, p. 88, \cite{COU} or even ref. \cite{R}). In such a case, one has the so-called Dirac's theory of second class constraints.
\end{itemize}

Let us compute the transverse Poisson structure\label{transverse Poisson} from our viewpoint, but supposing that {\itshape  the dimension of $M$, going through $z$, is the codimension in $W$ of the corresponding symplectic leaf ${\mathbf S}$ of the foliation} and so $$T_{{z}}W = T_{{z}}{M} \oplus T_{{z}}{\mathbf S},$$ (if $M$ is transversal to ${\mathbf S}$ so that $T_zM \cap \sharp_z\,\mathsf{Ann}\,T_zM \not= \{0\}$ is a distribution of constant rank then we shall need more assumptions -- see Vaisman \cite{IV}, for instance). 

Consider the following composition map 
$$
\sharp_{{\mathbf S}}\equiv {\sharp}\circ {\mathbf p}^{\star}_{T{\mathbf S}} \colon T^{\star}{\mathbf S}\to TW\vert_{\mathbf S}.
$$
If ${\mathbf p}^{\star}_{TM} = \mathbf{I} - {\mathbf p}^{\star}_{T{\mathbf S}}$ is the complementary projector and if we set 
\[
\sharp_{M}\equiv {\sharp}\circ {\mathbf p}^{\star}_{TM} \colon T^{\star}M\to TW\vert_{\mathbf S},
\]
then obviously $\sharp_{{\mathbf S}} + \sharp_{M} = \sharp$ and for all forms $\alpha, \beta$ one has
\[
\alpha \sharp_{{\mathbf S}} \beta + \alpha \sharp_{M} \beta  = \alpha \sharp \beta = {\boldsymbol \Pi}_{W}(\alpha,\beta).
\]

This expression suggests that $\alpha \sharp_{{\mathbf S}} \beta$ (resp. $\alpha \sharp_{M} \beta$) is a good candidate for the Poisson tensor ${\boldsymbol \Pi}_{\mathbf S}$ (resp. ${\boldsymbol \Pi}_M$). In fact, if we suppose that $\,\sharp \circ {\mathbf p}^{\star}_{T{\mathbf S}} = {\mathbf p}_{T{\mathbf S}} \circ \sharp\,$ then it is easy to show that ${\w}_{\mathbf S}\,(\sharp_{\mathbf S}(\alpha),\sharp_{{\mathbf S}}(\beta)) = \alpha \sharp_{{\mathbf S}} \beta$, where ${\w}_{\mathbf S}$ is the symplectic structure on ${\mathbf S}$, and so, from Proposition 3.2 of ref. \cite{R}, one has ${\boldsymbol \Pi}_{W}({\mathbf p}^{\star}_{T{\mathbf S}}\,(\alpha),{\mathbf p}^{\star}_{T{\mathbf S}}\,(\beta)) = \alpha \sharp_{{\mathbf S}}\beta$. 

Let us set 
${\boldsymbol \Pi}_{\mathbf S} = {\mathbf p}_{T{\mathbf S}}\,({\boldsymbol \Pi}_{W})$, that is,
$${\boldsymbol \Pi}_{\mathbf S}(\alpha,\beta)= {\boldsymbol \Pi}_{W}({\mathbf p}^{\star}_{T{\mathbf S}}\alpha, {\mathbf p}^{\star}_{T{\mathbf S}}\beta) =\alpha\,\sharp_{{\mathbf S}}\,\beta.
$$
Then 
$${\boldsymbol \Pi}_{M}={\mathbf p}_{TM}\,({\boldsymbol \Pi}_{W})=(\mathbf{I}-{\mathbf p}_{T{\mathbf S}})\,({\boldsymbol \Pi}_W)={\boldsymbol \Pi}_W-{\boldsymbol \Pi}_{\mathbf S}
$$
gives the complementary relation ${\boldsymbol \Pi}_{M}(\alpha, \beta) = \alpha \sharp_{M} \beta$.

The above tensor ${\boldsymbol \Pi}_M$ is a Poisson tensor, as a consequence of the symplectic structure of ${\mathbf S}$. So, let us see this in terms of local conditions related to the symplectic manifold ${\mathbf S}$: let $z_0 \in U$, an open subset of $W$ and $(x^1, \cdots, x^{2s})$ local coordinates for ${\mathbf S}$, used to define $$M\cap U = \{z \in U \subset W; x^a (z) = 0, a= 1, \cdots, 2s\},$$ transversal to ${\mathbf S}$, through $z_0$ . Then,
\begin{equation}
\label{PT}
{\w}_{\mathbf S}(X^a,X^b)= {\boldsymbol \Pi}_{\mathbf S}(d{x}^a,d{x}^b) = \{{x}^a,{x}^b\}_W=X^b({x}^a)= d{x}^a(X^b)=\lambda^{ab}.
\end{equation}
Here $X^a$ is the Hamiltonian vector field associated to $d{x}^a$, by the bundle homomorphism $\sharp_{\mathbf S}$ defined by ${\w}_{\mathbf S}$. For what follows, we shall denote by $\lambda_{ab}$ the entries of the inverse matrix of $(\lambda^{ab})$.  

Let $f \in {\mathcal C}^{\infty}(M)$ be such that $\{x^a,f\}_W=X^a(f)=df(X^a)=0$ and 
${\overline{f}}\in C^{\infty}(W)$ an extension of $f$ to a neighborhood in $W$, written as
\begin{equation}
\label{RR}
{\overline{f}}=f +{u}_b\,{x}^b,
\end{equation}
so that the action of $X^a$ on both sides of (\ref {RR}) gives ${u}_b = 
-X^a({\overline{f}})\lambda_{ab}$, and so 
\[
f={\overline{f}}+X^a({\overline{f}})\,\lambda_{ab}{x}^b = (I-{x}^b \,\lambda_{ba} \, X^a){\overline{f}}.
\] 

Now, the tensor field ${\tau}=(I-{x}^b\,\lambda_{ba}\,X^a)$ is a projector. To see this it is sufficient to show that
${\sigma}={x}^b\lambda_{ba}\,X^a$ is the complementary projector. Indeed, $\sigma(x^c) = x^c$ and as $\sigma(g) = X^a(g)\,\lambda_{ab}\,x^b$ then
\begin{eqnarray*}
\sigma^2(g) &=& X^a(g)\,\lambda_{ab}\,\sigma(x^b)\\
&=& X^a(g)\,[\lambda_{ab}\,\lambda^{bc}\,\lambda_{cd}\,x^d]= X^a(g)\,\lambda_{ab}\,\delta^b_d\, x^d\\
&=& X(g)\, \lambda_{ab}\,x^b = \sigma(g),
\end{eqnarray*}
and obviously ${\tau}({x}^a)=0$, $\tau \circ \sigma = \sigma \circ \tau =0.$
Thus the projector ${\mathbf p}^{\star}_{T{\mathbf S}}=T^{\star}{\mathbf S}\rightarrow T^{\star}W\vert_M$ is taken as ${\mathbf p}^{\star}_{T{\mathbf S}}=d{x}^a\,\lambda_{ab}\,X^b$ (or $\lambda_{ab}\,d{x}^a\otimes X^b$) and so from (\ref{PT}) we have, 
$$
{\mathbf p}^{\star}_{T{\mathbf S}}\,(d{x}^b)= d{x}^b\,(X^a)\, \lambda_{ac}\,d{x}^c = [\lambda^{ba}\,\lambda_{ac}]\,d{x}^c=d{x}^b
;\;\;{\mathbf p}^{\star}_{T{\mathbf S}}(df)= 0, 
$$
for the functions $f\in C^{\infty}(M)$ such that $X^a(f)=0$.
The dual projector ${\mathbf p}_{T{\mathbf S}} \colon TW\vert_M \to T{\mathbf S}$ is then given by 
\begin{equation}\label{DP}
{\mathbf p}_{T{\mathbf S}} = -X^a\,\lambda_{ba}d{x}^b,
\end{equation}
as $\w_{\mathbf S}$ is skew-symmetric. Then it follows that $${\mathbf p}_{T{\mathbf S}}(X^c) = -X^a\, \lambda_{ba}\,d{x}^b\,(X^c)=
X^c;\;\;\;{\mathbf p}_{T{\mathbf S}}(X_f)=0,$$
and so the composition $ \sharp_{{\mathbf S}} \equiv {\sharp}\,{\mathbf p}^{\star}_{T{\mathbf S}}\colon T^{\star}{\mathbf S}\to TW\vert_{\mathbf S}$ gives $\sharp_{{\mathbf S}}(d{x}^a)=X^a$. The Poisson structure ${\boldsymbol \Pi}_{{\mathbf S}}$ on $W$ is obtained from ${\w}_{\mathbf S}$, 
$$
{\boldsymbol \Pi}_{{\mathbf S}}= \frac{1}{2}\,\lambda_{ab}\,\sharp_{{\mathbf S}}(d{x}^a)\wedge
\sharp_{{\mathbf S}}(d{x}^b)=\frac{1}{2}\,\lambda_{ab}\,X^a\wedge X^b, 
$$
or directly from ${\boldsymbol \Pi}_{W}$ by projection: 
$${\boldsymbol \Pi}_{{\mathbf S}}(\alpha,\beta)= {\mathbf p}_{T{\mathbf S}}({\boldsymbol \Pi}_W )(\alpha,\beta)={\boldsymbol \Pi}_W({\mathbf p}^{\star}_{T{\mathbf S}}(\alpha),{\mathbf p}^{\star}_{T{\mathbf S}}(\beta)).$$

The complementary relation gives the transverse structure ${\boldsymbol \Pi}_{M}$:
\[
{\boldsymbol \Pi}_{M}= {\boldsymbol \Pi}_W - {\boldsymbol \Pi}_{{\mathbf S}}= {\boldsymbol \Pi}_W -\frac{1}{2}\, \lambda_{ab}\,X^a \wedge X^b.
\]
As 
\[
{\boldsymbol \Pi}_{\mathbf S}(d{\overline{f}},d{\overline{g}}) = \frac{1}{2}\, \lambda_{ab} \, X^a \wedge X^b \,(d{\overline{f}},d{\overline{g}}) = \{{\overline{f}}, x^a\}_W\, \lambda_{ab} \, \{{x}^b, {\overline{g}} \}_W,
\]
for extensions $\overline{f},\overline{g}$, if we set,
\[
\{f,g\}_M  = {\boldsymbol \Pi}_{W} (d{\overline{f}}, d{\overline{g}}) - {\boldsymbol \Pi}_{S}(d{\overline{f}},d{\overline{g}}),
\]
then it follows the well known Dirac's bracket formula:
\begin{equation}
\label{XZ}
\{f, g\}_M = \{{\overline{f}},{\overline{g}}\}_W - \{ {\overline{f}}, {x}^a\}_{W} \, \lambda_{ab}\, \{{x}^b,{\overline{g}}\}_{W}. 
\end{equation}
Note that, if we take into account (\ref{PT}) and (\ref{DP}), then \[
{\w}_{\mathbf S}({\mathbf p}_{T{\mathbf S}}(X_{{\overline{f}}}),{\mathbf p}_{T{\mathbf S}} (X_{{\overline{g}}})) = d{x}^c (X_{{\overline{f}}}) \, \lambda_{cd}\, d{x}^a \, (X_{{\overline{g}}}) \, \lambda_{ab}\, {\w}_{\mathbf S} \, (X^d,X^b)
\]
\[
 = \{{\overline{f}}, {x}^c\}_W \, \lambda_{ca} \{{x}^a,{\overline{g}}\}_W,
\]
as (obviously) it would be expected.
\section{Examples}
\subsection{The Chaplygin-Caratheodory's sleigh}\label{4.1}
We first recall that the Euler-La\-gran\-ge equation for non-holonomic mechanical systems, using the projector method, is implemented as follows.

Consider a mechanical system described by a Lagrangian $L$ defined on the bundle $TW$ of an $n$-dimensional configuration manifold $W$, $L(z^I,v^I,t)=T(z^I,v^I,t)-V(z^I,v^I)$, where the
$z^I$'s are the coordinates of $W$ and $v^I$'s are the velocities, submitted to non-holonomic constraints of the form \begin{equation}\label{01}
A^a_I(z^I)v^I + B^a(t)\,=\,0, \,\, 1\leq\,a\,\leq m .
\end{equation}

The stationarity condition of the action takes the well-known form
\begin{equation}
\label{w}
\int_{t_1}^{t_2} \left[\frac {d}{dt}\left(\frac{\partial L}{\partial v^I}\right)-\frac{\partial L}{\partial z^I}\right]{\delta z}^Idt\equiv \int_{t1}^{t_2} E_I\,{\delta z}^Idt=0,
\end{equation}

From (\ref{01}), it follows that the virtual displacements ${\delta z^I}$ are not all independent, since they must satisfies $m$ equations
\begin{equation}
\label{ff}
A^a_I(z^I)\,{\delta z^I}=0,
\end{equation}
such that all the  
$$
E_I=\frac {d}{dt}\left(\frac{\partial L}{\partial {v}^I}\right)-\frac{\partial L}{\partial {z}^I}
$$
cannot be set zero.

In order to set up the projector method it is more convenient to express all the equations in the matricial form. Thus (\ref {w}) and (\ref {ff}) are respectivelly written as

\begin{equation}
\label{dd}
\int_{t_1}^{t_2}{\bf E}^t\, {\delta {\bf z}}\,dt=0
\end{equation}
\begin{equation}
\label{bb}
{\bf A}\,{\delta {\bf z}}=0
\end{equation}
where  ${\bf E}$, ${\delta {\bf z}}$ are $n\times 1$ matrices, ${\bf A}$ is the $ m\times n$ matrix $(A^a_I)$ and the superscript t denotes transpose. 

Next we shall assume that the Riemannian metric ${\boldsymbol g}$ is given by the kinetic energy $T=({1}/{2})((v^1)^2 \cdots +(v^n)^2)$. We set ${\bf v}^{\star}={\boldsymbol g}^{\star}{(\bullet,\bf v)}$ for the associated co-vector. The constraint equation at the right of (\ref {bb}) splits the $TW$ (with respect to ${\boldsymbol g}$) such that $TW={\mathcal D} \oplus {\mathcal D}^{\bf c}$, where  ${\mathcal D}$ is the space of all virtual displacements compatible with the constraints. If ${\delta}\,{\eta}$ is an arbitrary (linearly independent) virtual displacement, then  the virtual displacement compatible with the constraints can be written as
\begin{equation}
\label{c}
{\delta {\bf z}}={\bf P}^t\,{\delta}\,{\eta}
\end{equation}
where ${\bf P}$ is the projector ${\bf P}\colon TW\rightarrow {\mathcal D}$. Substituition of (\ref {c}) in (\ref {dd}) gives 
$$
\int_{t_1}^{t_2}{\bf E}^t\,{\bf P}^t\,{\delta {\eta}}\,dt=0
$$

Now as ${\delta {\eta}}$ are linearly independent, we have
\begin{equation}
\label{ba}
{\bf P}\,{\bf E}=0,
\end{equation}
that is, the equation of motion compatible with the constraints must be such that, at each point, the Euler-Lagrange vector field is a vertical vector, while the virtual displacement ${\delta {\bf q}}$ is a horizontal vector, according to the D'Alembert Principle: ${\bf E}^t\,{\delta {\bf z}}=0$.

The equation (\ref {ba}) can be re-written as
\begin{equation}
\label{bi}
{\bf E}={\bf Q}\,{\bf E},
\end{equation}
where  ${\bf Q}\colon TW\rightarrow {\mathcal D}^{\bf c}$ is the complementary projector,  defined by
\begin{equation}
\label{rr}
{\bf Q}=({\bf A }^{\star})^t\,{\bf G^{-1}}\,{\bf A},
\end{equation}
where ${\bf G}={\bf A}\, ({\bf A }^{\star})^t$ is a non-sigular matrix for all mechanical systems known. Notice that the right-hand side of (\ref{bi}) gives the constraint forces.

To illustrate these features we apply the projector method to  obtain the equation of motion for the Chaplygin-Caratheodory's sleigh, a mechanical example of the Lagrange's Problem (we remit to Chaplygin, ref. \cite {Ch}, Carath\'eodore \cite{Ct}, Neimark \& Fufaev, ref. \cite{NFUV} and  the paper of Koiller, ref. \cite{Koiller0}, in particular \S\, $4.1$, for further details; for a general geometric setting on the subject see ref. \cite{Cantrijn3}).

The sleigh consists of a rigid body supported on a  horizontal plane, by three points, two of which slide freely (without friction)  and the third which is a knife edge (or the edge of a cutting wheel) rigidly fixed on the body.  We will consider here the special case studied by Caratheodory \cite {Ct,Rg}, in which the center of mass (c.m) lies  on the straight line, $\ell$, passing through the point of support, $p$, of the knife edge. The position of the body, in a fixed coordinate system in the horizontal plane, is determined by the coordinates $(x, y)\in \Ri^2$ of  $p$, and the angle ${\theta}$ between the line $\ell$ and the $x$-axis (and so the configuration manifold is $W = \Ri^2 \times S^1$). So, when the body slides, the velocity of $p=p(t)$ can be decomposed in a component along the  line $\ell$
\begin{equation}
\label{eq1Chp}
u= {\dot x}\,cos{\theta}+ {\dot y}\,sin{\theta}
\end{equation}
and a component perpendicular to it
\begin{equation}
\label{eq2Chp}
v=-r \,{\dot {\theta}} + {\dot y}\,cos{\theta}-{\dot x}\, sin{\theta}
\end{equation}
where  $r$ is the distance from the c.m to the point $p$ and $\dot{x}, \dot{y}, \dot{\theta}$ are the velocity coordinates (the dot indicate differentiation with respect to $t$). The non-holonomic constraint is given by the condition that the point $p$ can move freely on the plane only in the direction along the line $\ell$, but not in the direction perpendicular to it, that is 
\begin{equation}
\label{eq3Chp}
v=-r\,{\dot {\theta}} + {\dot y}\,cos{\theta}-{\dot x}\, sin{\theta}=0\, .
\end{equation}
This condition is expressed, by the  following restriction on the virtual displacements: 
\begin{equation}\label{const-1}
-r\,{\delta{\theta}} + cos{\theta}\,{\delta y}- sin{\theta}\,{\delta x}=0\, .
\end{equation}
In this case, we have
\begin{equation}\label{s}
{\bf A}=(-sin{\theta},cos{\theta},-r)\,.
\end{equation}

To symplify, we set the mass of the system equal to unity and also we shall restrict the example to the simple case where no external forces or torques exists: so
\begin{equation}
\label{d}
L=T=\frac{1}{2}((\dot{x})^2 + (\dot{y})^2 + {\mathsf J}\,\dot{\theta}^2),
\end{equation}
where ${\mathsf J}$ is the momentum of the inertia about the symmetry axis through the c.m. Therefore,  we obtain  ${\mathbf g}$ in matrix notation,
\begin{gather}\label{gmat}
\begin{pmatrix}
1& 0& 0\cr
0& 1& 0\cr
0& 0& {\textstyle{\mathsf J}}
\end{pmatrix}\, .
\end{gather}

Using (\ref {eq3Chp}), (\ref {s}) and (\ref {gmat}) we obtain
\begin{gather*}
{\mathbf Q}=\frac{{\mathsf J}}{{\mathsf J} + {r^2}}\,
\begin{pmatrix}
{\textrm{sin}^2\, \theta } & 
-\textrm{cos}\,\theta\,\textrm{sin}\,\theta
& r\,\textrm{sin}\,\theta \cr 
-\textrm{cos}\, \theta\,\textrm{sin}\,\theta
& \textrm{cos}^2\,\theta & 
-r\,\textrm{cos}\,\theta \cr 
{\mathsf J}^{-1}\,r\,\textrm{sin}\,\theta & 
- {\mathsf J}^{-1}\,r\,\textrm{cos}\,\theta & 
{\mathsf J}^{-1}\,r^2 
\end{pmatrix}
\,.
\end{gather*}

\begin{remark}
The action of the constraint form defined by (\ref{const-1}),  
$$\alpha=- \textrm{sin}\,\theta\,dx + \textrm{cos}\,\theta\,dy - r\,d\theta$$on the vector field
$$ X= - \textrm{sin}\,{\theta}\frac{\partial}{\partial x} + \textrm{cos}\,{\theta}\frac{\partial}{\partial y} -\frac{r}{{\mathsf J}}\frac{\partial}{\partial \theta}$$
gives 
$$
{\alpha}(X)=\frac{\textstyle{\mathsf J} + r^2}{\textstyle{\mathsf J}}\stackrel{\textrm{def}}{=} {G},
$$ and so ${\bf Q}={\mathbf g^{\star}}\,X\otimes{\alpha}$. We notice also that $||X||^2 = {\mathbf g}^{\star}(X,X)={G}$.$\quad \Box$
\end{remark}

To conclude this Lagrangian description, and to recover the equations found by Chaplygin and Caratheodory (\cite{Ct}, \cite{NFUV, Koiller0}) we first obtain the projected equations of motion taking into account (\ref{bi}): 
\begin{gather}\label{wa}
\begin{pmatrix}
\ddot{x} \cr \ddot{y}\cr \ddot{\theta}\cr
\end{pmatrix}=
\frac{{\mathsf J}}{{\mathsf J} + {r^2}}
\,
\begin{pmatrix}
{\textrm{sin}^2\, \theta } & 
-\textrm{cos}\,\theta\,\textrm{sin}\,\theta
& r\,\textrm{sin}\,\theta \cr 
-\textrm{cos}\, \theta\,\textrm{sin}\,\theta
& \textrm{cos}^2\,\theta & 
-r\,\textrm{cos}\,\theta \cr 
{\mathsf J}^{-1}\,r\,\textrm{sin}\,\theta & 
- {\mathsf J}^{-1}\,r\,\textrm{cos}\,\theta & 
{\mathsf J}^{-1}\,r^2 \cr
\end{pmatrix}\,
\begin{pmatrix}
\ddot{x} \cr \ddot{y}\cr \ddot{\theta}\cr
\end{pmatrix}\,.
\end{gather}

So, taking into account the constraint equation (\ref{const-1}), its first derivative, and  the system (\ref {wa}), we obtain the Chaplygin-Caratheodory equations
\begin{equation}\label{redeq11}
\dot{\omega} = - \bigg(\frac{r}{{\mathsf J} + {r^2}}\bigg)\,  \,u\,\omega, \quad {\dot u}=r\,{\omega}^2,
\end{equation}
where $ {\omega} = {\dot {\theta}}$.

The Hamiltonization is straightforwardly obtained, since the fiber derivative of $L$  $${\mathcal L}_L \colon TW \to T^{\star}W$$is a diffeomorphism. Locally, ${\mathcal L}_L (x,y,\theta,\dot x,\dot y,\dot \theta) = (x,y,\theta, p_x,p_y,p_{\theta}), p_I= (\partial L/\partial v^I)$, where the $v^{\prime}$s are the $\dot x,\dot y,\dot \theta$, and the $p^{\prime}$s are $(\partial L/\partial {\dot x})$, etc. Therefore, the constraint submanifold is defined by $- \textrm{sin}\,\theta\,p_1 + \textrm{cos}\,\theta\,p_2 - (r/{\mathsf J})\,p_{\theta} = 0$ 
and the Hamiltonian is $H= (1/2)(p^2_x +p^2_y + (1/{\mathsf J})\,p^2_{\theta}).$ Clearly, we reproduce the same arguments as above to obtain the constrained equations.

\subsection{A  particle subjected to the constraint $w=dz-ydx$.}
Let us consider a free particle in $M=\Ri^3$, with coordinates $(x,y,z)$, subjected to the non-holonomic contact form $w=dz-ydx$. The motion of the particle in the configuration space is given by the Lagrangian $L=({1}/{2})((v^x)^2 +(v^y)^2+(v^z)^2)$, where $(v^x,v^y,v^z)$ are the corresponding coordinates on the fibers of $TM$, and the Riemannian metric $\mathbf{g}$ on $M$ is given by $L$. So ${\mathbf g}$ is the identity tensor $dx\otimes dx + dy\otimes dy + dz\otimes dz$, and the metric expressed in the basis $\{dx,dy,w\}$ is
\[
{\mathbf g^{\star}}=(1+y^2)\,dx\otimes dx + y\,dx\otimes w + dy\otimes dy + y\, w\otimes dx + w\otimes w.
\]
Notice that (see \ref{met-ger-1}) $$g_{a \alpha} - g_{\alpha\beta}\,\Gamma^{\beta}_{a}\not=0.$$  With respect to the dual basis $\{X_1 = \partial/\partial x + y\, \partial/\partial z, X_2=\partial/\partial y, Y_3=\partial/\partial z\}$ one has, in matrix notation, 
\begin{gather*}
\mathbf g_{\star}=\begin{pmatrix}1&0&-y\cr
	     0&1&0\cr
	-y&0&1+y^2
\end{pmatrix}
\end{gather*}
and ${G}=(1+y^2)\,\partial/\partial z\otimes \partial/\partial z$ implies ${G}(w,w)=(1+y^2)\not=0$. Now,
\[
{\mathbf g_{\star}}(w,\bullet)= -y\, X_1 + (1+y^2)\, \frac{\partial}{\partial z} 
=\frac{\partial}{\partial z}-y\, \frac{\partial}{\partial x} = \xi,
\]
and so (\ref{proj-M}) is given by
\begin{eqnarray*}
{\mathbf Q}&=&{\mathbf g^{\star}}(\bullet,\xi)\,[{G}(w,w)]^{-1}\,{\mathbf g_{\star}}(w,\bullet)\\
&=&\frac{1}{1+y^2}\,\,w\otimes \xi\\
&=&\frac{1}{1+y^2}\,[y^2\,dx\otimes \frac{\partial}{\partial z} - y \,dx \otimes \frac{\partial}{\partial z} - y\,dz\otimes \frac{\partial}{\partial x} + dz\otimes \frac{\partial}{\partial z}]
\end{eqnarray*}
Therefore, in matrix notation, ${\mathbf Q}$ and ${\mathbf P}$ are
\begin{gather*}
{\bf Q}=\frac{1}{1+y^2}\begin{pmatrix}y^2 & 0 &-y\cr
0 & 0& 0\cr
-y & 0 &1
\end{pmatrix}
,\,\,
{\mathbf P}=\begin{pmatrix}
\frac{\textstyle{1}}{\textstyle{1+y^2}}&0&\frac{\textstyle{y}}{\textstyle{1+y^2}} \cr
0&1&0\cr
\frac{\textstyle{y}}{\textstyle{1+y^2}}&0&\frac{\textstyle{y^2}}{\textstyle{1+y^2}}
\end{pmatrix}
\end{gather*}

The ${\mathbf P}$-vectors are 
$$
X^{\mathbf P}_1=\frac{\partial }{\partial x}+y\frac{\partial }{\partial z};\;\;\;X^{\mathbf P}_2=\frac{\partial }{\partial y},
$$
and the ${\mathbf Q}$-vector $X^{\mathbf Q}$ is $\xi$. Again, remark that  $||X^{\mathbf Q}||=\sqrt{{\mathbf g^{\star}}(X^{\mathbf Q},X^{\mathbf Q})}=(1+y^2)$ and so we may also write ${\mathbf Q}=(||X^{\mathbf Q}||)^{-1}\,{\mathbf g^{\star}}(\bullet,X^{\mathbf Q})\,{\mathbf g_{\star}}(w,\bullet)$.

The relationship of the above ${\mathbf Q}$ with the orthogonal case, that is, $g_{a \alpha} - g_{\alpha\beta}\,\Gamma^{\beta}_{a}=0$ is the following. Let us denote by {\bf q} the projector ${\mathbf Q}$ for this new situation (see (\ref{defproj-1})). Then 
\begin{gather*}
{\bf q}=\begin{pmatrix}
0&0&0\cr
0 &0&0\cr
-y&0&1
\end{pmatrix}
\end{gather*}
As the ${\mathbf P}$-vector fields are in the kernel of ${\bf q}$, we deduce that $X^{\bf p}_1=X^{\mathbf P}_1\;\;X^{\bf p}_2=X^{\mathbf P}_2$, and $X^{\bf q}={\partial }/{\partial z}$.

It is easy to verify that $$[X^{\mathbf p}_1,X^{\mathbf p}_2]=X^{\mathbf q} ;\;\;[X^{\mathbf p}_1,X^{\mathbf q}]=0;\;\;\;[X^{\mathbf p}_2,X^{\mathbf q}]=0.$$ This is just the Heisenberg algebra, a very well known non-holonomic Lie algebra (and a fundamental example in sub-Riemannian geometry).  Also, 
\begin{gather*}
{\mathbf U}=\begin{pmatrix}
1+y^2&0&-y\cr
0 &1&0\cr
0&0&1
\end{pmatrix}
\end{gather*}
is so that ${\mathbf Q}={\mathbf U}\,{\boldsymbol q}\,{\mathbf U}^{-1}$. 

Let us now consider the following constraint equation 
\begin{equation}\label{const-2}
f(v^x,v^y,v^z)= v^z - y\,v^x = 0.
\end{equation}
for this problem. The use of (\ref {ba}), 
\begin{gather}
\label{UU}
\begin{pmatrix}
\frac{{\textstyle 1}}{{\textstyle 1+y^2}}&0&\frac{{\textstyle y}}{{\textstyle 1+y^2}} \cr
0 &1&0\cr
\frac{{\textstyle y}}{{\textstyle 1+y^2}}&0&\frac{{\textstyle y^2}}{{\textstyle 1+y^2}}
\end{pmatrix}
\begin{pmatrix}
{\ddot x}\cr
{\ddot y}\cr
{\ddot z}
\end{pmatrix}=\begin{pmatrix}
0\cr
0\cr
0\end{pmatrix}
\end{gather}
gives
$$
\frac{1}{1+y^2}\left({\ddot x}+y{\ddot z}\right)=0, \,\,\,
{\ddot y}=0. 
$$
Therefore,
\begin{equation}
{\ddot x}+y\,{\ddot z}=0, \,\,\,
{\ddot y}=0. \label{1}
\end{equation}

Equation (\ref {1}) is just the {\itshape momentum equation} of Bloch et Al. \cite{Bloch} (page $85$). We remark that from (\ref{const-2}) one obtains ${\ddot z}={\dot y}{\dot x}+y{\ddot x}$ and so, using this equation in   (\ref {1}) we obtain the Bates-Sniatycki motion's equations  (ref. \cite{Bat})
$${\ddot x}+\frac{y}{1+y^2}\,{\dot x}\,{\dot y}=0,\,\,\,
{\ddot y}=0.
$$

The motion of the particle in the phase space (endowed with the natural Poisson structure $\boldsymbol \Pi$) is given by the Hamiltonian
$$H=\frac{1}{2}\left(p_x^2+p_y^2+p_z^2\right),$$
restricted to the submanifold defined by the equation $p_z-yp_x=0$. 
The momenta, compatible with the constraint, can be written as ${\overline {\mathbf p}}={\mathbf P}\,{\mathbf p}$ or ${\overline p_i}={\mathbf P}_i^jp_j$, where $i,j=x,y,z$. Thus,
\begin{gather}
\label{U}
\begin{pmatrix}
{\overline p_x}\cr
{\overline p_y}\cr
{\overline p_z}
\end{pmatrix}=\begin{pmatrix}
\frac{\textstyle{1}}{\textstyle{1+y^2}}&0&\frac{\textstyle{y}}{\textstyle{1+y^2}} \cr
0&1&0\cr
\frac{\textstyle{y}}{\textstyle{1+y^2}}&0&\frac{\textstyle{y^2}}{\textstyle{1+y^2}}
\end{pmatrix}\,\begin{pmatrix}
p_x\cr
p_y\cr
p_z
\end{pmatrix},
\end{gather}
Note that $ {\overline p_z}=\frac{y}{1+y^2}\left(p_x+yp_z\right)\equiv
y{\overline p_x}$. The application of the local definition of the Poisson brackets on a regular Poisson manifold gives
$$ \{z^i,z^j\}=0;\;\;\;\{z^i,{\overline p_j}\}={\bf P}^i_j;\;\;\;\{{\overline
p_i},{\overline p_j}\}=\left({\bf P}^k_j\frac{\partial {\bf
P}_i^l}{\partial z^k}-   {\bf P}^k_i\frac{\partial {\bf
P}^l_j}{\partial z^k}\right)p_l,$$
and so, after using  the constraint equation $p_z=yp_x$, we obtain the following {\itshape pseudo-Poisson structure}, that is, the Poisson bracket is skew-symmetric, satisfies de Leibniz rule but it may not satisfy de Jacobi identity (see Marle \cite {CMM}, for further details):
\begin{gather}
\label{V}
{\overline {\boldsymbol \Pi}}=\begin{pmatrix}
0&0&0&\frac{\textstyle{1}}{\textstyle{1+y^2}}&0\cr
0&0&0&0&1\cr
0&0&0&\frac{\textstyle{y}}{\textstyle{1+y^2}}&0\cr
-\frac{\textstyle{1}}{\textstyle{1+y^2}}&0&-\frac{\textstyle{y}}{\textstyle{1+y^2}}&0&-\frac{\textstyle{y\,p_x}}{\textstyle{1+y^2}}\cr
0&-1&0&\frac{\textstyle{y\,p_x}}{\textstyle{1+y^2}}&0\cr
-\frac{\textstyle{y}}{\textstyle{1+y^2}}&0&-\frac{\textstyle{y^2}}{\textstyle{1+y^2}}&0&\frac{\textstyle{p_x}}{\textstyle{1+y^2}}
\end{pmatrix}.
\end{gather}

Finally, we want to remark that there are many others non-holonomic systems which admits a pseudo - Poisson structure. The Chaplygin-Caratheodory's sleigh (example $4.1$) is the protoptype of mechanical systems described by the Poisson geometry.

In fact, most of the known  non-holonomic mechanical systems are given by completely non-integrable constraints which are linear in the velocities (as in the examples $4.1$ and $4.2$). In these cases, as we have shown in $4.2$, the induced Poisson structure is given by:
\begin{gather}
\label{R1}
{\overline {\boldsymbol \Pi}}=\begin{pmatrix}
{\mathbf 0}&{\mathbf P}\cr
-{\mathbf P}&{\mathbf D}
\end{pmatrix}.
\end{gather}
 where ${\mathbf P}$ is the $n\times n$ matrix-projector and ${\mathbf D}$ is the $n\times n$ matrix whose elements are 
$${\mathbf D}_{ij}=\left({\mathbf P}^k_j\frac{\partial {\mathbf P}_i^\ell}{\partial z^k}- {\mathbf P}^k_i\frac{\partial {\mathbf P}^\ell_j}{\partial z^k}\right)p_\ell$$ 

Generally, the matrix ${\overline {\boldsymbol \Pi}}$ is singular because the matrix ${\mathbf P}$ is generally singular, that is det\, ${\mathbf P}=0$ . We say "generally" because one could conceive cases in which the system of constraints is completely non-integrable but - nevertheless - admits a singular integral which reduces the system to the integrable form. In this case, as in the holonomic case, ${\overline {\boldsymbol \Pi}}$, is  invertible, which gives rise a symplectic  structure. The authors do not know any concrete examples of non-holonomic mechanical systems of this kind.


\begin{thebibliography}{99}
\bibitem{ArnoldMec} {Arnol'd, V. I.}, 1984, {\itshape Mathematical Methods of Classical Mechanics}, Springer Verlag, Berlin.
\bibitem{Arnoldo} {Arnol'd, V. I., Kozlov, V. V. , Neishtadt, A. I.}, 1988, {\itshape Dynamical Systems III}, Encyclopedia of Mathematical Sciences, Vol. 3, Springer-Verlag, N.Y./Berlin.
\bibitem{Bloch} {Bloch, A. M., Krishnaprasad, P. S., Marsden, J. E., Murray, R. M.}, 1996, {\itshape Non-holonomic Mechanical Systems with Symmetry}, Arch. Rational Mech. Anal, {\bf 136}, 21-99.
\bibitem{Bat} {Bates, L., Sniatycki, J.}, 1992, 
{\itshape Non-holonomic reduction}, Rep. Math. Phys., {\bf 32}, 99-115.  
\bibitem{Bel} {Bela\"{\i}che, A., Risler, J-J.}, 1996, {\itshape Sub-Riemannian Geometry}, Birkh\"auser, Basel.  
\bibitem{Brezinski} {Brezinski, C.}, 1997,  {\itshape Projection methods for systems of equations}, Elsevier, Amsterdam.
\bibitem{Cantrijn3}  {Cantrijn, F., Cort\'es, J., de Le\'on, M., Mart\'{\i}n de Diego, D.}, 2002, {\itshape On the geometry of generalized Chaplygin systems}, Math. Proc. Cambridge Phil. Soc, {\bf 132}, 323-351.
\bibitem{Cantrijn}  {Cantrijn, F., de Le\'on, M., Mart\'{\i}n de Diego, D.}, 1999, {\itshape On almost-Poisson structures in non-holonomic mechanics I}, Nonlinearity, {\bf 12}, 3, 721-737.
\bibitem{Cantrijn2}  {Cantrijn, F., de Le\'on, M., Mart\'{\i}n de Diego, D.}, 2000, {\itshape On almost-Poisson structures in non-holonomic mechanics II: the time-dependent framework}, Nonlinearity, {\bf 13}, 4, 1379-1409.
\bibitem{Ct} {Caratheodory, C.}, 1933, {\itshape Der Schlitten}, Z.Angew. Math. Mech. 13, 71-76.
\bibitem{Ca} {Caratheodory, C.}, 1991, {\itshape Calculus of variations and Partial Differential Equations of the First Order}, Third Edition, A.M.S. 
\bibitem{Cardin}  {Cardin, F., Favretti, M.}, 1996, {\itshape On non-holonomic and vakanomic dynamics of mechanical systems with non-integrable constraints}, J. Geom. Phys., {\bf 18}, 295-325.
\bibitem{Cartan}  {Cartan, E.}, 1928, {\itshape Sur la repres\'entation g\'eom\'etrique des syst\`emes mat\'eriels non holonomes}, Proc. Int. Congr. Math., Bologna, {\bf 4}, 253-261.
\bibitem {Ch} Chaplygin, S.A., 1911, {\itshape On the theory of motion of nonholonomic systems. Theorems on the reducing multiplier}. Mat. Sb 28, 303-314. 
\bibitem {COU} Courant, T. J., 1990, {\it Dirac Manifolds}, Trans. Amer. Math. Soc., {\bf 319}, 2,  631-660.
\bibitem{CMDM}  {Cort\'es, J., de Le\'on, M., Mart\'{\i}n de Diego, D., Mart\'{\i}nez, S.}, {\itshape Geometric description of vakonomic and nonholonomic dynamics. Comparision of solutions}. To appeaer in SIAM Journal of Control and Optimization.
\bibitem{LMurray}  {Lewis, A. D., Murray, R. M.}, 1995, {\itshape Variational principles for constrained systems: theory and experiments}, International Journal of Nonlinear Mechanics, {\bf 30}, 6, 793-815.
\bibitem {Da} {Dazord, P.}, 1992, {\itshape M\'ecanique hamiltonienne en pr\'esence de constraintes}. Institut de Math\'ematique et Informatique, Universit\'e Claude Bernard-Lyon I. Preprint URA CNRS 746.
\bibitem{LichneM} {Dazord, P., Lichnerowicz, A., Marle, C.M.}, 1991, {\itshape Structure locale des  vari\'et\'es de Jacobi}, J. Math. pures et appl., {\bf 70}, 101-152.
\bibitem{Rg} Gran-Olasson, R., 1933, {\itshape Some remarks on a paper by Caratheodory}, Z.Angew. Math. Mech. 13, 71 -76) (see also 1954, Proc.Internat. Congr. Math.2, Noordhoff, Groningen and North-Holland, p. 349). 
\bibitem{Hermann} {Hermann, R.}, 1992, {\itshape Constrained Mechanics and Lie Theory}, Math Sci. Press, Massachusetts.
\bibitem{Koiller0} {Koiller, J.}, {\itshape Reduction of some classical non-holonomic systems with symmetry}, Arch. Rational Mech. Anal., {\bf 118}, 1992, 113-148.
\bibitem{KRP} {Koiller, J., Rodrigues, P. R., Pitanga, P.}, 2001, {\itshape Nonholonomic connections following \'Elie Cartan}, An. Acad. Bras. Cienc, {\bf 73}, 2, 165-190.
\bibitem{KRP1} {Koiller, J., Rodrigues, P. R., Pitanga, P.}, 2001, {\itshape Sub-Riemannian Geometry and Non-holonomic Mechanics}, Contemporary Mathematics, {\bf 288}, 353-357.
\bibitem{Kupka0} {Kupka, I.}, 1996, {\itshape G\'eom\'etrie sous-riemannienne}, S\'eminaire Bourbaki, {\bf 817}, 351-380.
\bibitem{Kupka1} {Kupka, I., Oliva, W.}, 2001, 
{\itshape Non-holonomic Mechanics}, J. Diff. Equations, {\bf 169}, 169-189.
\bibitem{deLR-0} {de Le\'on, M., de Diego, D. M.}, 1996, {\itshape On the geometry of non-holonomic Lagrangian systems}, J. Math. Phys., {\bf 37}, 7, 3389-3414.
\bibitem{deLR-01} {de Le\'on, M., de Diego, D. M.}, 1996, {\itshape Solving non-holonomic Lagrangian dynamics in terms of almost product structures}, Extracta Mathematicae, {\bf 11}, 2, 325-347.
\bibitem{deLR} {de Le\'on, M.,  Rodrigues, P. R.}, 1989, {\itshape Methods of Differential Geometry in Analytical Mechanics}, North Holland Mathematical Studies, {\bf 158}, Elsevier, Amsterdam.
\bibitem{CMM} {Marle, C-M.}, 1997, {\itshape Various approaches to conservative and non-conservative non-holonomic systems}. Proc. Workshop on non-holonomic constraints in dynamics (Calgary, August 26-29, 1997), Rep. Math. Phys., {\bf 42}, 211-29.
\bibitem{JMR} Marsden, J. E., and Ratiu, T., 1986, {\it Reduction of Poisson Manifolds}, Lett. Math. Phys, {\bf 11}, 161-169. 
\bibitem{MR1} {Marsden, J. E., Ratiu, T.}, 1994, {\itshape Introduction to Mechanics and symmetry}, Springer Verlag.
\bibitem{Montbook}  {Montgomery, R.}, 2002, {\itshape A Tour of Subriemannian Geometries, Their Geodesics and Applications}, AMS Surveys and Monographs, {\bf 91}.
\bibitem{NFUV} {Neimark, J., and Fufaev, N.}, 1972, {\itshape Dynamics of Nonholonomic Systems}, Amer. Math. Society, Providence, R. I.
\bibitem{Nickerson} {Nickerson, H. K., Spencer D. C.}, 1955, {\itshape Differentiable manifolds and sheaves}, Princeton Univ..
\bibitem {OH} {Oh., Y. G.}, 1986, {\itshape Some Remarks on the Transverse Poisson Structures of Coadjoint Orbits}, Lett. Math. Phys. {\bf 12}, 87-91.
\bibitem{PITA} {Pitanga, P.}, 1994, {\itshape Projector method and constrained systems}, N. Cimento, {\bf 109 B},2, 113-119.
\bibitem{R} {Pitanga, P., Rodrigues, P. R.}, 1997, {\itshape On the projector method for regular Poisson manifolds}, J. Phys A: Math. Gen., {\bf 30}, 8719-8725.
\bibitem{Poor} {Poor, W. A.}, 1981, {\itshape Differential geometric structures}, Mc-Graw Hill, N. Y.
\bibitem{Rei} {Reinhart, B. L.}, 1959, {\itshape Foliated Manifolds with bundle-like metrics}, Annals of Mathematics, {\bf 69}, 1, 119-132.
\bibitem{Schouten} {Schouten, J. A.}, 1954, {\itshape Ricci Calculus}, Springer Verlag, Berlin.
\bibitem{Sommerfeld} Sommerfeld, A., 1952, {\itshape Mechanics},  Academic Press, USA. 
\bibitem{robotics} {Sussman, G. J., Wisdom, J.}, {\itshape Structure and Interpretation of Classical Mechanics}, 2001, The MIT Press, Cambridge, London.
\bibitem {IV} {Vaisman, I.}, 1994, {\itshape Lectures on the Geometry of Poisson Manifolds}, Progress in Mathematics, {\bf 118}, Birkhauser Verlag.
\bibitem{IV-0} {Vaisman, I.}, 1971, {\itshape Vari\'et\'es riemanniennes feuillet\'ees}, Czechosl. Math. J., {\bf 21}, 46-75.
\bibitem{IV-1} {Vaisman, I.}, 1973, {\itshape Cohomology and differential forms}, M. Dekker Inc., N.Y.
\bibitem{Schaft} {van der Schaft, A. J., Maschke, B. M.}, 1994, {\itshape On the Hamiltonian Formulation of Nonholonomic Mechanical Systems}, Rep. Math. Phys., {\bf 34}, 225-233.
\bibitem{vershik} {Vershik, A. M., Gershkovich, V. Y.}, 1990, {\itshape Nonholonomic Dynamical Systems, Geometry of Distribuitions and Variational Problems}, Dynamical Systems VII, Encyclopedia of Mathematical Sciences, Vol. VII, Springer-Verlag, N.Y./Berlin.
\bibitem{vershik1} {Vershik, A. M., Fadeev, L. D.}, 1981, {\itshape Lagrangian Mechanics in invariant form}, Selecta Math. Sov., {\bf 1}, 339-350.
\bibitem{Vilms} {Vilms, J.}, 1967, {\itshape Connections on tangent bundles}, J. Diff. Geometry, {\bf 1}, 235-243.
\bibitem{Walker} {Walker, A. G.}, 1955, 1958, {\itshape Connections for parallel distributions in the large I, II}, Quart. J. Math., Oxford, {\bf 2}, 6, 301-308; {\bf 9}, 221-231.
\bibitem {W} {Weinstein, A.}, 1983, {\itshape The Local Structure of Poisson Manifolds},  J. Diff. Geometry {\bf 18}, 523-557.
\end{thebibliography}
\end{document}